# The influence of quantum noise on the Grover algorithm and quantum Fourier transform: quantum operations theory approach

Yu.I. Bogdanov[1,2,3*], A.Yu. Chernyavskiy[1,4], B.I. Bantysh[1,3], D.V. Fastovets[1,3**], V.F. Likichev[1]

1 - Institute of Physics and Technology, Russian Academy of Sciences, 117218, Moscow, Russia
2 - National Research Nuclear University 'MEPHI', 115409, Moscow, Russia
3 - National Research University of Electronic Technology MIET, 124498, Moscow, Russia
4 - M. V. Lomonosov Moscow State University, 119991, Moscow, Russia

The method of noisy multiqubit quantum circuits modeling is proposed. The analytical formulas for the dependence of quantum algorithms accuracy on qubits count and noise level are obtained for Grover algorithm and quantum Fourier transform. It is shown that the proposed approach is very much in line with results obtained by Monte Carlo statistical modeling method.

The developed theory makes it possible to predict the influence of quantum noise on the accuracy of the forward-looking multiqubit quantum systems under development.

## INTRODUCTION

It is well known that the dynamics analysis of multi-particle quantum system is a difficult problem. More generally, there is no effective way to solve this problem because the dimension has exponential growth as the number of particles is increased. At the same time, the situation is aggravated if realistic open systems are being studied. It is very important to take into account the quantum states decoherence processes in such systems. Such consideration is necessary for realization of quantum technologies including perspective quantum computing devices. It is important to note that these devises allow us not only to solve different mathematical problems, but also to simulate large quantum mechanical systems efficiently [1-3]. This concept is a basic idea of quantum computer independently proposed by Yu. I. Manin [4, 5] and R. Feynman [6].

Actually, the quantum circuits model is a universal model of multi-particle quantum systems dynamics. This model determines the most common approach to quantum computations [7]. The systems and corresponding circuits are considered with strictly analytical approach if they consist of one or two particles. In other cases, numerical modeling is used. It has been studied extensively [8-11] due to the significance of such modeling for the development of quantum computers. However, all these articles have to address the above-mentioned problem associated with exponential growth of complexity. At present, the record number of simulated qubits is 45 [12] (in Russia – 39 qubits [13]). Furthermore, the implementation of quantum circuits simulation for distributed supercomputer systems leads to some serious difficulties [13]. Practically, these difficulties and dimension growth make the modeling of even 100 qubits impossible. Moreover, to simulate a system with consideration of quantum noise, we need to reduce the qubits number by half (in general quantum circuit analysis, without initial state definition, the number of qubits is reduced four-fold ).

* bogdanov_yurii@inbox.ru

** fast93@mail.ru

The new calculation method is proposed in this work. This method makes it possible to evaluate the accuracy of large dimensional quantum-mechanical circuits effectively. Sometimes, the developed approach makes it possible to obtain highly precise analytical estimates of the required values. Generally, this approach allows for analyzing large circuits with size inaccessible for direct modeling.

This article consists of two parts. There is a short introduction to quantum open systems formalism in the first part. The proposed method of high-dimensional quantum schemes accuracy calculation is described in the second part. Firstly, the developed method was applied to the Grover's algorithm - one of the most important quantum algorithms. Then, the improvement of this method was used to analyze another important quantum procedure – quantum Fourier transform.

## 1. NOISY QUANTUM CIRCUITS FORMALISM

Technically, arbitrary quantum circuit is defined by a set of unitary operations $\{U_i\}$. Operations $U_i$ usually affect one, two or more rarely three qubits. Implementation of each transform $U_i$ is imply the existence of Hamiltonian $H_i$ for real physical systems. This Hamiltonian performs the operation during time $t_i : U_i = e^{-iH_i t_i}$. However, we need only operators $U_i$ for the further consideration. The quantum circuits modeling problem consists of the computing of the output state $U_s \cdot U_{s-1} \cdot \ldots \cdot U_2 \cdot U_1 |\psi_0\rangle$ according to the initial state $|\psi_0\rangle$.

To describe open quantum systems affected by quantum noise, it is necessary to use the density matrix formalism. In this case, the quantum states transformations expand from unitary ones to so-called quantum operations [14, 15].

Arbitrary quantum operation can be represented by the operator-sum form:

$$\varepsilon(\rho) = \sum_i E_i \rho E_i^\dagger,$$

where operators $E_i$ are called Kraus operators and satisfy the normalization condition $\sum_i E_i^\dagger E_i = I$.

In the vector state terms, noise can be added by Monte Carlo approach. For example, arbitrary quantum gate $U_{ideal}$ can be replaced by its noisy analogue

$$U = U_{ideal} \cdot V_{noise}, \quad V_{noise} = \begin{pmatrix} \cos(e\xi) & \sin(e\xi) \\ -\sin(e\xi) & \cos(e\xi) \end{pmatrix}, \tag{1}$$

where $\xi$ is a random variable with normal distribution $N(0,1)$, $e$ is an error rate.

Such representation of a noisy transform is very valuable and will be used further.

The noisy controlled phase shift operator, which is used in quantum Fourier transform, can be represented similarly:

$$U_\theta = U_\theta^{ideal} \cdot V_\theta^{noise} = \begin{pmatrix} 1 & 0 & 0 & 0 \\ 0 & 1 & 0 & 0 \\ 0 & 0 & 1 & 0 \\ 0 & 0 & 0 & e^{-i(\theta + e \cdot \xi)} \end{pmatrix}. \quad (2)$$

Where $U_\theta^{ideal}$ - ideal controlled phase shift operator and $V_\theta^{noise}$ - corresponding noise operator.

An alternative instrument of quantum noise analysis is the Choi-Jamiolkowski isomorphism [14, 15]. An arbitrary quantum operation can be represented by the corresponding density matrix of a larger dimension. It is a basic idea of this isomorphism. To construct the Choi-Jamiolkowski relative state for the $d$-dimensional operation we need to take maximally-entangled state

$$\frac{1}{\sqrt{d}} \sum_{i=1}^{d} |i\rangle \otimes |i\rangle$$

and apply the operation to the second subsystem:

$$\rho_\varepsilon = \frac{1}{\sqrt{d}} \sum_{i,j=1}^{d} |i\rangle\langle j| \otimes \mathcal{E}(|i\rangle\langle j|).$$

Because of Choi-Jamiolkowski isomorphism we can analyze the quantum circuits without the reference to particular input states. At the same time, we consider the fidelity (as a circuit accuracy measure) between relative states of noisy and ideal circuits.

An important consequence of the Choi-Jamiolkowski isomorphism is the statistical equivalence of noise described by the Monte Carlo method (by unitary operators distributions) and Kraus operators, and a relatively simple transition from the first description to the second one.

It can be shown, that single-qubit noisy transform $V_{noise}$ (1) is statistically equivalent to the following set of Kraus operators

$$E_1 = \sqrt{\lambda_1}\begin{pmatrix} 1 & 0 \\ 0 & 1 \end{pmatrix}, \quad E_2 = \sqrt{\lambda_2}\begin{pmatrix} 0 & 1 \\ -1 & 0 \end{pmatrix},$$

where

$$\lambda_1 = \frac{1}{2}\left(1 + \exp(-2e^2)\right), \quad \lambda_2 = \frac{1}{2}\left(1 - \exp(-2e^2)\right). \quad (3)$$

Similarly, for the noisy controlled phase shift operator $V_\theta^{noise}$ from (2), we have

$$E_1 = \begin{pmatrix} 1 & 0 & 0 & 0 \\ 0 & 1 & 0 & 0 \\ 0 & 0 & 1 & 0 \\ 0 & 0 & 0 & \sqrt{P} \end{pmatrix}, \quad E_2 = \begin{pmatrix} 0 & 0 & 0 & 0 \\ 0 & 0 & 0 & 0 \\ 0 & 0 & 0 & 0 \\ 0 & 0 & 0 & \sqrt{1-P} \end{pmatrix}, \quad (4)$$

where $P = \exp(-e^2)$.

## 2. HIGH PRECISION FIXED RANK APPROXIMATION OF NOISY DYNAMICS IN GROVER'S ALGORITHM AND QUANTUM FOURIER TRANSFORM

Let the initial input sate of the circuit be pure. Then, the corresponding density matrix has unit rank. Each Kraus operator determines one of the quantum system evolution alternative ways. Therefore, each quantum operation increases the rank of the density matrix in geometric progression right up to the maximal possible rank $r = 2^n$, where $n$ is the number of qubits. To reduce the number of necessary computational resources, we will approximate the obtained density matrix of Choi-Jamiolkowski state by a lower-rank matrix at each step (the fixed amount of maximal eigenvalues are being retained, other are nullified). Such approximation preserves the positivity of density matrices. However, its trace decreases, and it corresponds to the proportion of representatives evolving along the «correct» path.

Fig. 1 shows the accuracy of such approximation for the calculation of fidelity during the Grover's algorithm for different values of the fixed rank.

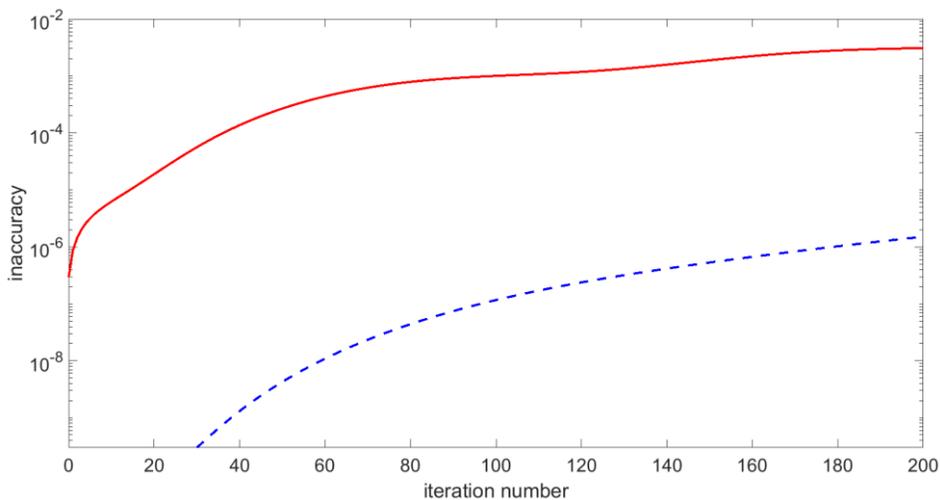

Fig. 1. The approximation error of 12 qubits noisy Grover's algorithm for rank 1 (solid line) and 30 (dashed line) respectively; $e = 0.01$ is the error rate.

The results of simulation show that rank-1 approximation ($r = 1$) provides a simulation accuracy significantly better than 1%. Moreover, the approximation accuracy increases with the number of qubits. The obtained result shows that all useful information about «correct» ideal quantum system evolution path is almost completely contained in the first (main) component only. The other components mostly contain noise and no useful information.

Rank-1 approximation leads to the following analytical accuracy estimate of the Grover's algorithm under the influence of noise:

$$p_{noise}(j) = \lambda_1^{n+2nj} p_{ideal}, \tag{5}$$

where $j$ is the step of the algorithm, $p_{noise}$ and $p_{ideal}$ are the probabilities of obtaining the correct answer (fidelity) by the noisy and ideal algorithm respectively, $n$ is the number of qubits, $\lambda_1$ is defined by (3). The $n$ in the power corresponds to the noise of initial state preparation and $2nj$ corresponds to the noise to two Hadamard gates on each algorithm step and qubit. It is important to note, that obtained estimate is a lower bound. In other words, the true estimate of correct answer probability will be slightly higher than the presented one (according to the content of useful information in the removed components).

Similar direct application of rank-1 approximation to quantum Fourier transform leads to the following estimate, which determines the transform accuracy of random state:

$$F_{QFT} = P_H^n P_R^{n(n-1)/8}, \qquad (6)$$

where $P_H = \frac{1}{2}(1 + \exp(-2e^2))$, $P_R = \exp(-e^2)$ are the correct answer result probabilities of Hadamard transform and controlled phase-shift gate respectively.

The presented formula uses the fact that Fourier transform contains $n$ Hadamard gates and $n(n-1)/2$ two-qubits phase transforms. The two-qubit noise generates a power $n(n-1)/8$ because the phase transform acts only on a quarter of amplitudes on each step for random input state.

The presented formula gives the lower-bound estimate of fidelity (solid curve on Fig. 2).

This estimate can be improved by *main components reduction* of Kraus operators (4) of the phase-shift gate. It is well known that Kraus operators are ambiguous with respect to some unitary transformation.

It is possible to select a unitary transform that makes the diagonals of considered Kraus operators orthogonal to each other. Kraus operators in new representation have form

$$\tilde{E}_1 = \frac{1}{\sqrt{1+f^2}} \begin{pmatrix} 1 & 0 & 0 & 0 \\ 0 & 1 & 0 & 0 \\ 0 & 0 & 1 & 0 \\ 0 & 0 & 0 & \sqrt{P} + f\sqrt{1-P} \end{pmatrix}, \quad \tilde{E}_2 = \frac{1}{\sqrt{1+f^2}} \begin{pmatrix} -f & 0 & 0 & 0 \\ 0 & -f & 0 & 0 \\ 0 & 0 & -f & 0 \\ 0 & 0 & 0 & -f\sqrt{P} + \sqrt{1-P} \end{pmatrix}, \qquad (7)$$

where $P = \exp(-e^2)$, $f = \dfrac{\sqrt{1+3P} - P - 1}{\sqrt{P(1-P)}}$.

The Kraus operators (7) and (4) are equivalent, because they determine the same Choi-Jamiolkowski state and the same quantum operation.

A refined fidelity lower bound for Fourier transform of random state is:

$$F_{QFT} = P_H^n \tilde{P}_R^{n(n-1)/8}, \qquad (8)$$

where $P_H = \lambda_1$, $\tilde{P}_R = \dfrac{\left(\sqrt{P} + f\sqrt{1-P}\right)^2}{\left(1+f^2\right)^4}$.

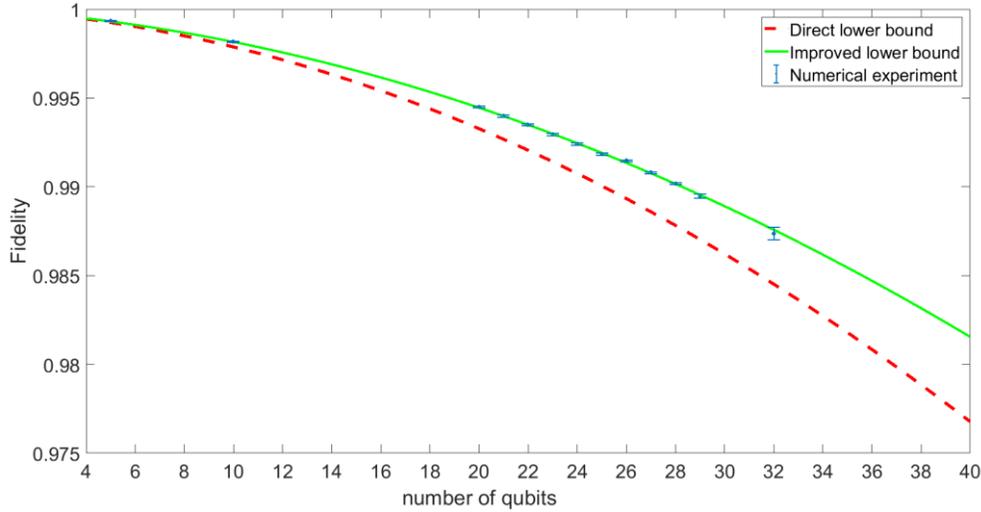

Fig. 2. The fidelity of quantum Fourier transform ($e = 0.01$) obtained by Monte Carlo method, straight rank-1 approximation and rank-1 approximation with main components reduction.

The Fig. 2 shows that the improved estimate has much better precision and visually coincides with the numerical Monte Carlo experiment.

The developed theory makes it possible to predict the quantum noise influence on accuracy of multi-qubit quantum systems. Such quantum systems have not been implemented in hardware yet and they cannot be modeled on any modern or perspective classical computers. This fact is illustrated on Fig.3.

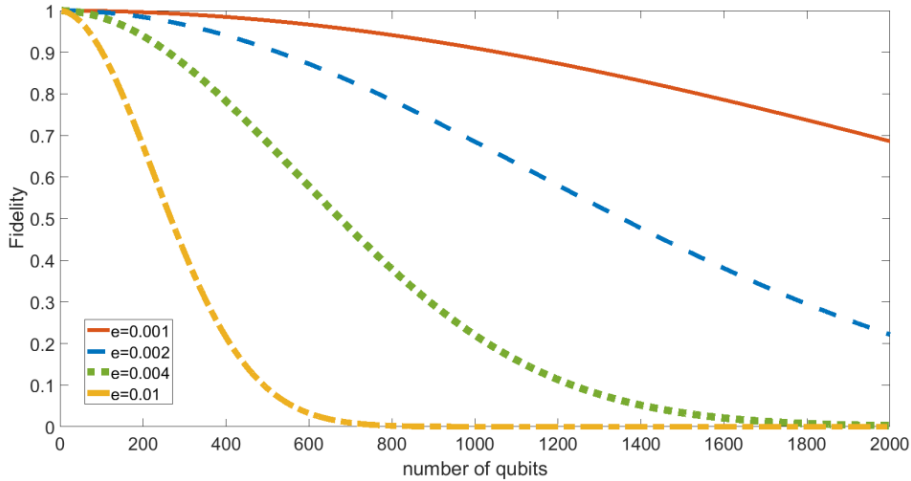

Fig. 3. Dependence of quantum Fourier transform accuracy on number of qubits for different noise rate.

These calculations provide the necessary noise level estimates for the quantum Fourier transform effective implementation. It is very clear that correct action of the experimental gates will be possible for sufficiently low values of noise level $e$. For example, noise level $e = 0.001$ provides 69% of quantum Fourier transform precision for 2000 qubits register. Changing this parameter by one order leads to the rapid quantum circuit accuracy degradation for large qubits number.

# CONCLUSION

Let us formulate the main results of this work.

The method of multi-qubit quantum systems modeling subject to the decoherence and quantum noise is developed. This method is based on quantum operations theory with fixed rank approximation.

High precise analytical estimates for Grover's algorithm and quantum Fourier transform are obtained. These estimates give the lower bounds of probabilities of algorithms correct answers. The dependences of these estimates on the number of qubits and noise rate are presented.

The proposed approach is aimed at the prediction of accuracy characteristics of the promising multi-qubit quantum systems under development.

The work was supported by Program of the Russian Academy of Sciences in fundamental research.